# Proposal and design of a new SiC-emitter lateral NPM Schottky collector bipolar transistor on SOI for VLSI applications

M. Jagadesh Kumar and D.V. Rao

**Abstract:** A novel bipolar transistor structure, namely, a SiC emitter lateral NPM Schottky collector bipolar transistor (SCBT) with a silicon-on-insulator (SOI) substrate is explored using two-dimensional (2-D) simulation. A comprehensive comparison of the proposed structure with its equivalent Si lateral NPN BJT and an SiC emitter lateral NPN HBT is presented. Based on simulation results, the authors demonstrate for the first time that the proposed SiC emitter lateral NPM transistor shows superior performance in terms of high current gain and cut-off frequency, reduced collector resistance, negligible reverse recovery time and suppressed Kirk effect over its equivalent Si lateral NPN BJT and SiC emitter lateral NPN HBT. A simple fabrication process compatible with BiCMOS technology is also discussed.

Silicon carbide (SiC) has become a very important material in the recent past because of its high thermal conductivity, high saturated electron drift velocity, high cut-off frequency, and its ability to operate in hostile and high temperature environments [1–3]. Moreover, the fabrication compatibility with silicon not only reduces the cost but also increases the yield. The above advantages of SiC, coupled with the advent of high quality local epitaxy and lateral epitaxial overgrowth of SiC [4–7], opened up new opportunities for device designers to use SiC as an emitter of BJT to improve device performance through the bandgap engineering mechanism. The SiC emitter HBTs [8, 9], which have been reported in the literature, are vertical in structure and based on bulk technology. Hence, they did not gain much popularity for VLSI applications. To take full advantage of SOI technology and the lateral Schottky collector transistor, along with the bandgap engineering mechanism of the SiC emitter, for the first time, we propose a new SiC emitter lateral NPM Schottky collector bipolar transistor for obtaining improved electrical characteristics without sacrificing the lithographic limits.

Based on two-dimensional simulation results, we demonstrate that the SiC emitter lateral NPM HBT shows better electrical characteristics in terms of high current gain, high cut-off frequency, complete elimination of Kirk effect, and approximately zero base storage time, compared with its equivalent Si lateral NPN BJT and SiC emitter lateral NPN HBT. In the following Sections steady-state, dynamic and transfer characteristics, and a possible fabrication process compatible with BiCMOS technology, are presented.

## 2 Device structure and parameters

Figure 1 shows the top and cross-sectional view of the wide bandgap SiC emitter lateral NPM Schottky collector bipolar transistor, which has been implemented in the two-dimensional device simulator ATLAS [10]. Emitter length is 3.8 um with doping equal to $5 \times 10^{19}$cm$^{-3}$ and base length is 0.4 um with p-type doping equal to $5 \times 10^{17}$cm$^{-3}$. SOI thickness is chosen to be 0.2 um and buried oxide thickness is 0.38 um. The emitter region is a wide bandgap SiC n-type material that can be formed by well-established deposition processes [5–7]. The base contact is obtained using the p$^+$-poly deposited on the p-type silicon base region. The Schottky contact is taken at the right edge of the base that acts as a metal collector. Aluminium is chosen as the metal collector since it gives a barrier height of 0.91 eV as reported in literature [11], based on experimental results. Aluminium also offers high conductivity and better process selectivity. The SiC emitter lateral NPN HBT, which has been used for comparison, has exactly the same dimensions and impurity concentrations as that of the SiC emitter lateral NPM HBT except that the collector doping of lateral NPN HBT is chosen to be 3x10$^{17}$cm$^{-3}$ so that both the devices have identical collector breakdown voltage $BV_{CEO}$ for zero base current. Similarly, the Si lateral NPN BJT has exactly the same dimensions as the SiC emitter structures except that the emitter region is $n^+$-silicon and its fabrication procedure is the same as reported in [12].

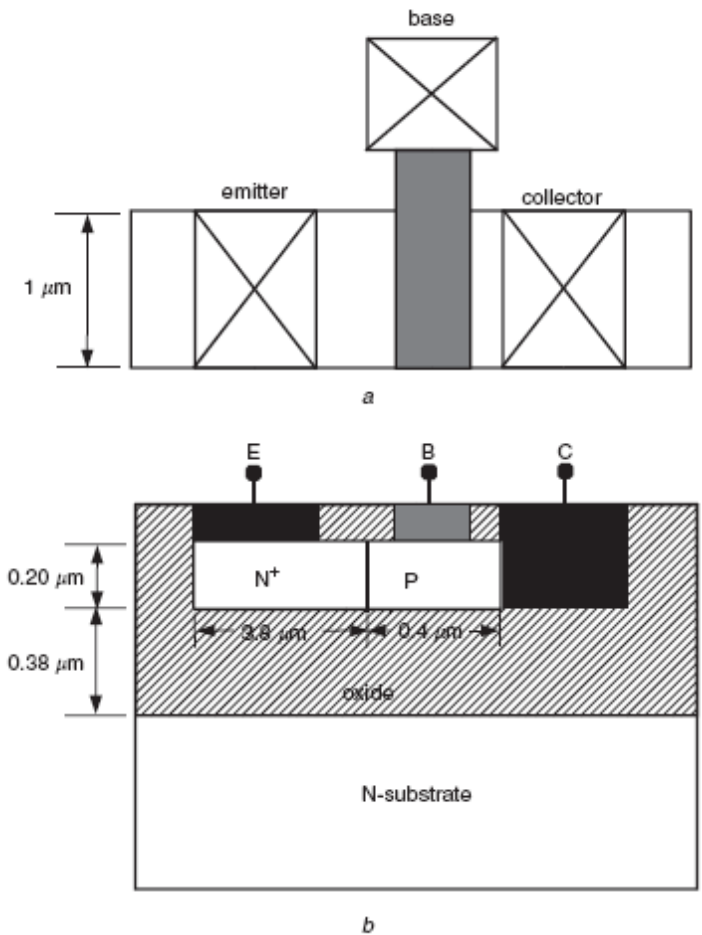

**Fig. 1** *SiC-emitter lateral NPM HBT implemented in this investigation*
*a* Top layout
*b* Cross-sectional view

**3 Proposed fabrication of SiC-emitter lateral NPM HBT**

The proposed fabrication process of the SiC emitter lateral NPM HBT is similar to the fabrication of SOI lateral NPN bipolar transistors [12] with a little modification. The modification involves epitaxial growth of the SiC emitter region and formation of the Schottky collector. We can begin with an SOI wafer having an epitaxial layer of thickness 0.2 um and doping of $5\_10^{17}cm^{-3}$. After mesa-isolation, a thick CVD oxide is deposited and patterned as shown in Fig. 2a. We then deposit the in situ n+ SiC on the vertical edge (at point X in Fig. 2b) of the silicon surface, which acts as a seed and the SiC grows laterally [4–7] as shown in Fig. 2b. After performing the CMP, we deposit a thick CVD oxide and pattern it as shown in Fig. 2c. Following this, a nitride film is deposited (Fig. 2d) and an unmasked RIE etch is performed until the planar silicon nitride is etched retaining the nitride spacer at the vertical edge of thick CVD oxide (Fig. 2e). Following this, a thick CVD oxide is deposited (Fig. 2f) and CMP process is carried out to planarise the surface. Selective etching is used to remove the nitride spacer, which will create a window in the oxide as shown in Fig. 2g. After depositing in situ p$^+$-poly into the opened window, the wafer is once again planarised using CMP leaving p$^+$-poly in the place where the nitride film was present (Fig. 2h). Now a mask is used for etching both the field oxide and the silicon film used to open a contact window for the Schottky metal collector as shown in Fig. 2i. Using another mask, the n$^+$ emitter contact window is opened by etching the field oxide (Fig. 2j). Following this, aluminium is deposited and patterned to form the Schottky collector and ohmic contacts on the emitter and p$^+$-poly base region. The final structure is as shown in Fig. 1b.

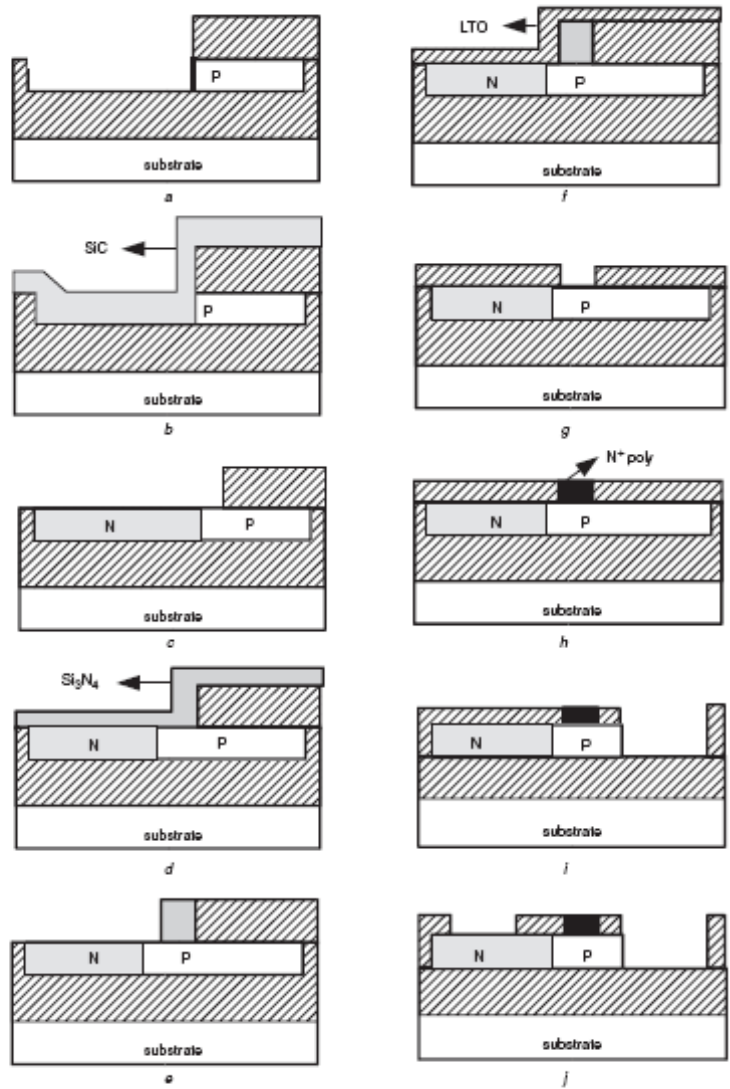

**Fig. 2** *Proposal process flow for lateral SiC-emitter NPM HBT*

## 4 Simulation results and discussion

To investigate and predict the theoretical performance of the SiC emitter lateral NPM transistor, a physically based numerical device simulator ATLAS [11] is used. It calculates the electrical characteristics, which are associated with specified physical structures and bias conditions. It provides the internal device behaviour by solving the well-established drift-diffusion transport equations. In our simulations we used the appropriate physical models such as concentration dependent mobility, field-dependent mobility, and Klassens mobility models and the bandgap narrowing effect is also taken into account. The Fermi–Dirac distribution is defined to calculate the carrier statistics, and Shockley–Read–Hall and Auger recombination mechanisms are also invoked in the simulation. The SiC material properties used for the simulation purpose are taken from the published results [13]. The incomplete ionisation model is included to consider the deep donor ($E_D$) and deep acceptor ($E_A$) levels in SiC emitter [13]. To account for the Schottky junction property, the standard thermionic emission model is specified, integrating the effect of image force barrier lowering phenomenon [14]. Numerical small signal analysis predicts the dynamic response. The simulated steady state and dynamic characteristics of the proposed lateral NPM HBT structure and its comparison with the equivalent Si lateral BJT and SiC emitter lateral NPN HBT structure are discussed below.

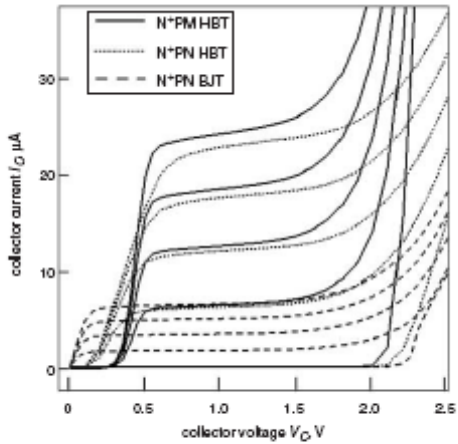

Fig. 3  *Common-emitter I–V characteristics of Si lateral NPN BJT, SiC-emitter lateral NPM and NPN HBTs*
Curves are for $I_B = 0$ to 200 nA, with intervals of 50 nA

*4.1 DC characteristics*

The simulated DC output characteristics of all the three lateral transistors are illustrated in Fig. 3. The Si lateral NPN BJT has a low current driving capability compared with the proposed lateral SiC emitter NPM HBT, which exhibits an enhanced collector current for identical base current. It may be pointed out that the SiC emitter lateral NPM structure shows a finite offset voltage of $V_{EC} = 0.3V$, which is typical of any Schottky collector transistor [15] and should be considered while designing the digital logic circuits. Figure 4 shows the Gummel plots for the Si lateral NPN BJT, SiC emitter lateral NPM and NPN HBTs for a fixed collector base voltage ($V_{CB} = 1V$). We observe that the SiC emitter NPM HBT exhibits a base current lower than that of the SiC emitter lateral NPN HBT due to a finite current flow from metal into the n-base when the Schottky collector junction is reverse biased [16]. It is also seen that the base current of the SiC emitter lateral NPM HBT is less than that of the SiC emitter lateral NPN HBT even at high-level injection of carriers, proving the absence of the Kirk effect [17]. However, in the case of the Si lateral NPN BJT, the increase in base current points to a significant base widening at high collector currents. The series collector resistance is governed by the doping concentration and carrier mobility in the collector drift region. The NPM structure with a metal as its collector has a low resistivity as compared to the n- type drift collector region of the SiC emitter lateral NPN HBT or Si lateral NPN BJT. This makes the Schottky collector structure immune to the base widening even at high collector currents. The reduced base current, along with better efficiency of minority carrier collection at the metal collector–base junction, gives rise to a higher current gain in the case of NPM HBT as compared with both the Si lateral NPN BJT and also the SiC emitter lateral NPN HBT as shown in Fig. 5.

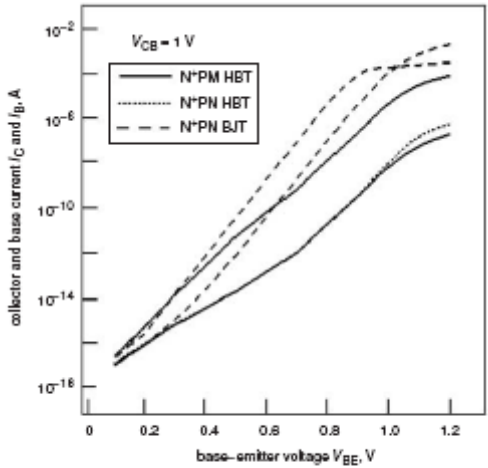

Fig. 4 *Gummel plots of Si lateral NPN BJT, SiC-emitter lateral NPM and NPN HBTs*

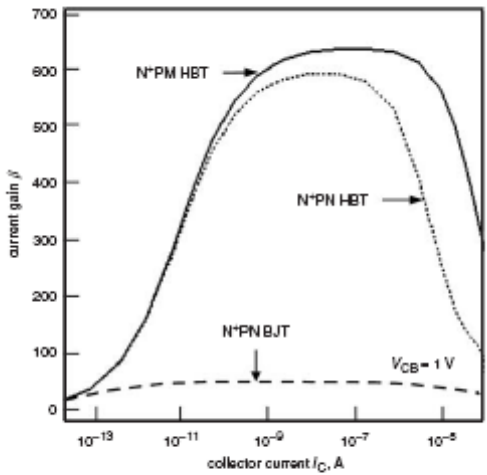

Fig. 5 *Beta against collector current of Si lateral NPN BJT, SiC-emitter lateral NPM and NPN HBTs*

### 4.2 Dynamic analysis
The simulated unity current gain cut-off frequency ($f_T$) against collector current is shown in Fig. 6 for the Si lateral NPN BJT, SiC emitter lateral NPM and NPN HBTs. The SiC-emitter lateral NPM HBT exhibits a higher cut-off frequency, since it offers the least collector resistance and better transconductance than the NPN HBT. At a collector current of 0.1mA, the $f_T$ is observed to be 5.2GHz, while for the comparable SiC-emitter lateral NPN HBT or Si lateral NPN BJT, there is a rapid fall in $f_T$ for the same current, due to the decrease in transconductance and base width widening. The transient behaviour of both SiC emitter lateral NPM and NPN HBT is shown in Fig. 7. It is clear from the Figure that the SiC-emitter lateral NPM HBT has approximately zero base charge storage time because of the absence of base widening and a negligible minority carrier lifetime in the metal collector region. However, the SiC-emitter lateral NPN HBT shows a higher storage time. Figure 8 illustrates the transfer characteristics of the inverter using the SiC emitter lateral NPM and NPN HBT. We observe that the SiC emitter lateral NPM HBT exhibits a better performance compared to its equivalent NPN HBT in terms of the steep (abrupt) transition region.

### 4.3 Temperature analysis
The temperature dependence of current gain against collector current of SiC emitter lateral NPM and NPN HBT is shown in Fig. 9. It is clear from Figure that current gain decreases with an increase in temperature. However, the SiC-emitter lateral NPM transistor still has a large current gain (285) even at 400 K, while NPN transistor has a current gain of 213 at 400K. This indicates that the SiC emitter lateral NPM transistor can be operated at higher ambient temperatures without a significant loss in current gain, while the current gain of Si lateral NPN BJT is far less than this value even at room temperature. This demonstrates the advantage of the proposed SiC-emitter lateral NPM structure for high temperature applications.

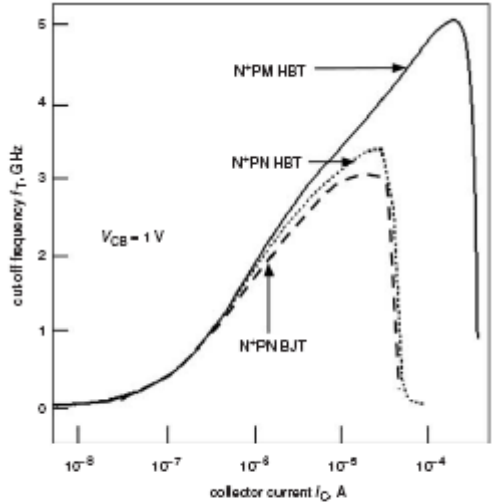

**Fig. 6** *Cut-off frequency against collector current of Si lateral NPN BJT, SiC-emitter lateral NPM and NPN HBTs*

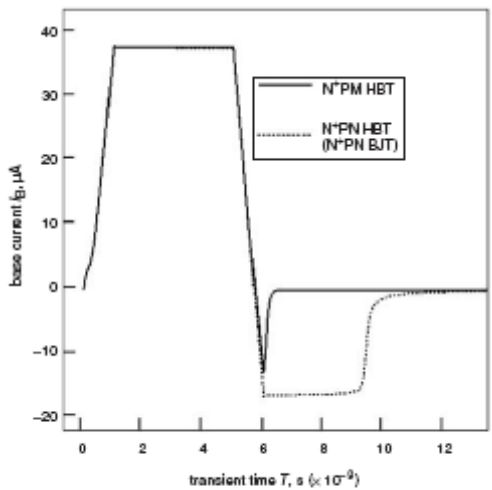

**Fig. 7** *Switching performance of Si lateral NPN BJT, SiC-emitter lateral NPM and NPN HBTs*

## 5 Conclusions

A comprehensive comparison of SiC emitter lateral NPM transistor is carried out with its equivalent Si lateral NPN BJT and SiC emitter lateral NPN HBT to investigate the performance improvements that can be realised using a wide bandgap emitter. Based on our simulation results, we arrive at the conclusion that the SiC emitter NPM HBT shows better characteristics in terms of higher current gain, higher cutoff frequency, negligible reverse recovery time and suppressed base width widening compared to the equivalent lateral SiC NPN HBT. The proposed structure may be the best candidate for high current-driving applications such as high speed DAC/ADC converters. Negligible reverse recovery time of NPM HBT not only improves response time but also reduces the power dissipation during the switching activities; thus it minimizes the power–delay product of the circuit under consideration.

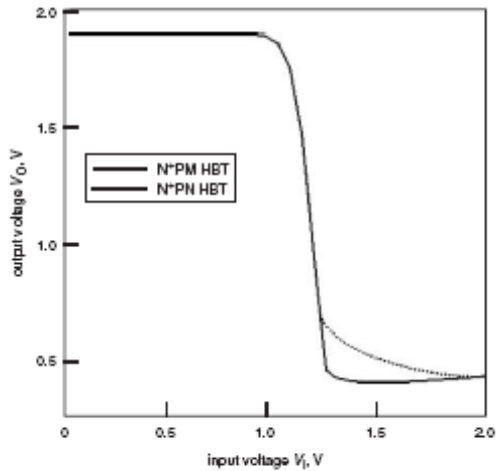

**Fig. 8** *Voltage transfer characteristics of SiC-emitter lateral NPM and NPN HBT inverter*

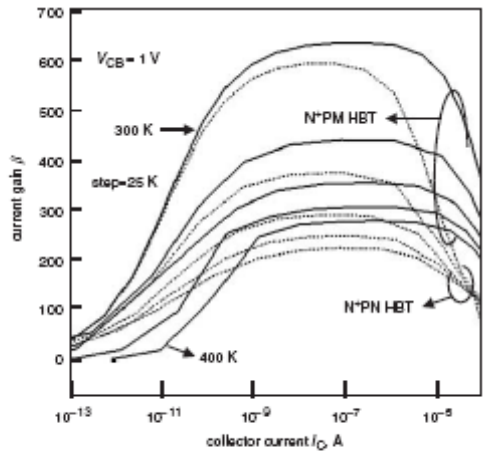

**Fig. 9** *Beta against collector current of SiC-emitter lateral NPM and NPN HBT for temperature range between 300 K and 400 K*